# Tuning intrinsic anomalous Hall effect from large to zero in two ferromagnetic states of SmMn$_2$Ge$_2$

Mahima Singh,[*] Jyotirmoy Sau,[*] Banik Rai,[*] Arunanshu Panda, Manoranjan Kumar,[‡] and Nitesh Kumar[‡]

*S. N. Bose National Centre for Basic Sciences, Salt Lake City, Kolkata 700106, India*

**ABSTRACT**

The intrinsic anomalous Hall conductivity (AHC) in a ferromagnetic metal is completely determined by its band structure. Since the spin orientation direction is an important band–structure tuning parameter, it is highly desirable to study the anomalous Hall effect in a system with multiple spin reorientation transitions. We study a layered tetragonal room temperature ferromagnet SmMn$_2$Ge$_2$, which gives us the opportunity to measure magnetotransport properties where the long *c*-axis and the short *a*-axis can both be magnetically easy axes depending on the temperature range we choose. We show a moderately large fully intrinsic AHC up to room temperature when the crystal is magnetized along the *c*-axis. Interestingly, the AHC can be tuned to completely extrinsic with extremely large values when the crystal is magnetized along the *a*-axis, regardless of whether the *a*-axis is magnetically easy or hard axis. First-principles calculations show that nodal line states originate from Mn-*d* orbitals just below the Fermi energy ($E_F$) in the electronic band structure when the spins are oriented along the *c*-axis. Intrinsic AHC originates from the Berry curvature effect of the gapped nodal lines in the presence of spin-orbit coupling. AHC almost disappears when the spins are aligned along the *a*-axis because the nodal line states shift above $E_F$ and become unoccupied. Since the AHC can be tuned from fully extrinsic to intrinsic even at 300 K, SmMn$_2$Ge$_2$ becomes a potential candidate for room-temperature spintronics applications.

---

[‡] manoranjan.kumar@bose.res.in
[‡] nitesh.kumar@bose.res.in
[*] These authors contributed equally

**INTRODUCTION**

Topology in time-reversal symmetry-broken systems has recently been the subject of intense research. Berry curvature effects, which are at the center of these topological materials, can manifest themselves in momentum space or real space to impart exotic transport properties [1–3]. The Berry curvature induced anomalous Hall effect is one such avenue that has recently engaged solid-state physicists and materials scientists [4,5]. In addition to the obvious van der Waals ferromagnets, crystal structure features such as Kagome and square net subunits have attracted attention to layered ferromagnets [5–8]. An important feature of these layered ferromagnets is that they often exhibit strong uniaxial magnetic anisotropy. Many recent studies indicate large anomalous Hall conductivity in layered ferromagnets when the magnetic field is applied along the easy magnetic axis [6,8,9]. Often, it is not possible to access different crystallographic axes as easy axes of magnetization in a structurally anisotropic compound without structural phase transition. For example, the long crystallographic *c*-axis remains the easy magnetization axis for many hexagonal and tetragonal systems [9,10]. Therefore, it is highly desirable to study systems in which one has the freedom to study the anomalous Hall effect (AHE) by tuning the easy magnetic axis from one crystallographic direction to another. Another important feature in the band structure that enhances the effect of Berry curvature for the anomalous Hall effect is nodal line states [4,9,11]. These states are generally present in systems with high crystallographic symmetry containing multiple mirror planes [12,13].

We chose $SmMn_2Ge_2$ to study the anisotropic magnetotransport property because it is known to exist in at least three magnetically ordered states with spontaneous alignment of spins along in-plane and out-of-plane directions [14]. In the $RMn_2Ge_2$ family of compounds, where *R* corresponds to the rare-earth elements, the magnetic states are known to be dictated by the intralayer closest Mn-Mn distance "*d*" within the Mn square nets [15,16]. It is known from the literature that $d < 2.87$Å favors antiferromagnetic ordering, while $d > 2.87$Å favors ferromagnetic ordering [14]. $SmMn_2Ge_2$ stands out for its proximity to the critical value of *d*. Temperature is the obvious parameter that can smoothly change *d* across the critical value. Indeed, temperature leads to multiple magnetic phase transitions and reentrant ferromagnetism. $SmMn_2Ge_2$ attains ferromagnetism at $T_C \sim 350$ K. As the temperature decreases from $T_C$, the ferromagnetic (FM) phase undergoes a sharp transition to an antiferromagnetic (AFM) phase around 150 K. This is because the thermal contraction causes *d* to decrease, eventually falling below the critical value of 2.87 Å [17,18]. In the AFM state, it is suggested that the interaction between Sm and Mn atoms weakens compared to the coupling between Mn-Mn atoms.

However, as the temperature drops below 100 K, the Sm-Sm coupling becomes intense and is sufficient to break the AFM Mn-Mn coupling, resulting in the emergence of the reentrant ferromagnetic (RFM) phase in the system [14].

Some sister compounds to SmMn$_2$Ge$_2$ have recently gathered attention due to the occurrence of a large topological Hall effect [19–21]. In this work, we have carried out a detailed study of the magnetotransport properties of SmMn$_2$Ge$_2$ single crystals across various phases along with first-principles calculations. We show that the nature of the anomalous Hall effect can be transformed from completely intrinsic to extrinsic depending on the direction of the applied magnetic field. By performing angle-dependent magnetotransport studies, we explain the topological Hall-like feature by considering magnetic domain dynamics. First-principles calculations demonstrate that the intrinsic anomalous Hall conductivity is a result of the nodal line states just below the Fermi energy, which depends strongly on the direction of spin alignment.

**RESULTS**

SmMn$_2$Ge$_2$ crystals grow in a tetragonal ThCr$_2$Si$_2$-type crystal structure with space group *I*4/*mmm*. Within this structure, all Sm, Mn, and Ge atoms are stacked in layers along the *c*-axis occupying 2*a* (0,0,0), 4*d* (0,0.5,0.25), and 4*e*(0,0,*z*) Wyckoff positions, respectively, as shown in Fig. 1(a) [22,23]. Mn atoms form a square-net structure in the *ab*-plane, which is represented by a yellow plane as a guide to the eye. An optical image of a typical crystal is shown in the inset of Fig. 1(b). The composition of the crystals is very close to SmMn$_2$Ge$_2$ as is evident from the energy-dispersive x-ray spectroscopy (EDX) data presented in Fig. S1(a). The XRD pattern [see Fig. 1(b)] obtained on a crystal by keeping the flat plane along the sample holder reveals reflections corresponding only to (00*l*) planes, confirming that the exposed plane is the *ab*-plane. To confirm the single-crystal nature of the sample, we performed Laue x-ray diffraction by exposing the x-ray perpendicular to the flat plane. The diffraction pattern consists of fourfold symmetric points, which is consistent with a tetragonal structure with space group *I*4/*mmm*. Sharp spots in the pattern indicate good quality of the single crystal. The diffraction pattern fits well with the *I*4/*mmm* space group using the reported cell parameters [see Fig. S1(b)].

Figure 1(d) shows the magnetization (*M*) versus temperature (*T*) data obtained by applying a magnetic field of 200 Oe along the *c*-axis. As mentioned above, with decreasing temperature, the paramagnetic SmMn$_2$Ge$_2$ becomes ferromagnetic at ~340 K, where the moment on the Mn and Sm layers points out of the layer (along the *c*-axis). We call this FM phase FM-1. As the temperature is decreased, an AFM phase is achieved at ~140K, which can be inferred from the vanishing net magnetic moment. The AFM persists up to ~100 K, below which it attains a reentrant FM phase, which we call the FM-2 phase. The FM-2 phase is stable down to the lowest measurement temperature of 2 K. The spins on Mn and Sm orient themselves ferromagnetically in the *ab*-plane. This is also evident from the seven times smaller net magnetization of the FM-2 phase compared to that in the FM-1 phase. We have shown the *M* versus *T* data when the magnetic field of 200 Oe is applied in the *ab*-plane along [100] in the Fig. S2. All the features are the same except that the FM-2 phase has much higher magnetization than the FM-1 phase. The important observation here is that the *c*-axis is the easy axis of magnetization in the FM-1 phase, which changes to the easy *ab*-plane of magnetization in the FM-2 phase. Longitudinal resistivity ($\rho_{xx}$) as a function of temperature when current is applied along the *a*-axis decreases with decreasing temperature, confirming the metallic nature of the material. The slope of the $\rho_{xx}$ versus *T* data changes at the paramagnetic to FM-1 phase transition, and we also observe small kinks corresponding to the FM-1 to AFM and AFM to FM-2 phase transitions, which are consistent with previous reports. The resistivity at 300 K and 2 K are 228 μΩ-cm and 1.55 μΩ-cm, respectively, giving a large residual resistivity ratio $\left(RRR = \frac{\rho_{xx}^{300K}}{\rho_{xx}^{2K}}\right)$ of 147. For some other crystals (see Fig. S3) we found an even larger RRR of 278 demonstrating the excellent quality of the grown crystals.

In Fig. 2(a) we show $M - \mu_0 H$ data while applying the magnetic field along the *c*-axis at various temperatures across the FM-1, AFM and FM-2 phases. In the FM-1 phase (see 300 K curve), the magnetization saturates at a small field of 0.16 T in accordance with the spontaneous alignment of the spins along the *c*-axis. The saturation magnetization ($M_S$) is ~2 $\mu_B$/f.u. In the FM-2 phase (see 2 K curve), the magnetization increases continuously and does not saturate up to 5 T because the *c*-axis is the hard axis in this phase. In the AFM phase (see 125 K curve), we observe a spin-flop transition at 1 T where there is a sudden increase of magnetization followed by a slow saturation of magnetization at ~1.75 T to a value of ~3.1$\mu_B$/f.u. When the magnetic field is applied along [100] as shown in Fig. 2(b), in the FM-1 phase (see 300 K curve) the magnetization is difficult to saturate (3 T, to a value of ~2 $\mu_B$/f.u.) compared to the FM-2 phase (see 2 K curve) where it easily saturates to a value of ~3.7 $\mu_B$/f.u. at a small magnetic field of 0.4 T. In the AFM phase (see 125 K curve) we observe a spin-flip transition at ~0.4 T, corresponding to a sharp saturation of the magnetization. These observations are in good agreement with the previous reports [14].

To investigate the effect of magnetic anisotropy on the electrical transport behavior of SmMn$_2$Ge$_2$ in different crystallographic directions, we performed magnetic field dependent measurements of $\rho_{xx}$ and $\rho_{yx}$. The graphs depicted in Fig. 3(a) and (b) illustrate the magnetoresistance $[MR = \frac{\rho_{xx}(H) - \rho_{xx}(0)}{\rho_{xx}(0)} \times 100\ \%]$ of SmMn$_2$Ge$_2$ at different temperatures, with the magnetic field applied along [001] and [100], respectively. We encounter both positive and negative MR at various temperature and magnetic field ranges. We attribute the positive *MR* to the orbital effect, whereas the negative *MR* can be attributed to the decrement of scattering due to spin alignment.

Figure 4(a) shows the Hall resistivity ($\rho_{yx}$) with the current flowing along [100] and the magnetic field applied along [001] at temperatures corresponding to the FM-1 phase. In metallic systems with spontaneous magnetization, the total Hall resistivity can be expressed as: $\rho_{yx} = \rho_{yx}^N + \rho_{yx}^A = R_0 B + R_S M_S$ where, $\rho_{yx}^N$ is the normal Hall resistivity component, $\rho_{yx}^A$ is the anomalous Hall resistivity component, $R_0$ is the normal Hall coefficient, $R_S$ is the anomalous Hall coefficient and $M_S$ is the saturation magnetization. At 300 K, $\rho_{yx}$ increases rapidly before saturating at ~0.16 T and then changes very slowly with field. The behavior of $\rho_{yx}$ mimics that of *M*, suggesting the presence of an AHE. An additional small slope in the $\rho_{yx} - \mu_0 H$ data in the saturation region provides $R_0$. Anomalous Hall resistivity ($\rho_{yx}^A$) is defined as the finite Hall resistivity at zero magnetic field. However, due to the absence of the coercivity in the $\rho_{yx} - \mu_0 H$ data, the value of the anomalous Hall resistivity ($\rho_{yx}^A$) is estimated by fitting the normal Hall region with a linear function and extending it to zero field to obtain $\rho_{yx}^A$ as the intercept on the *y*-axis (see Fig. 4(a)). The Hall conductivity $\sigma_{xy}$ is calculated from the tensor relation $\sigma_{xy} = \frac{\rho_{yx}}{\rho_{yx}^2 + \rho_{xx}^2}$. The same protocol as for $\rho_{yx}^A$ is used to estimate the value of $\sigma_{xy}^A$ from the magnetic field dependent data of $\sigma_{xy}$. $\sigma_{xy}^A$ is known to arise mainly through three main mechanisms, namely the intrinsic Karplus-Luttinger mechanism, and the extrinsic skew scattering and side jump mechanisms [2]. Since the intrinsic contribution to $\sigma_{xy}^A$ is entirely electronic Berry curvature-dependent and does not depend on the scattering effects, it is expected to be temperature-independent. Figure 4(b) shows the temperature dependence of $\sigma_{xy}^A$ in the FM-1 phase. It has a weak temperature dependence with the lowest and highest values being ~105 Ω$^{-1}$cm$^{-1}$ and 77 Ω$^{-1}$cm$^{-1}$, respectively, accounting for about a 25% decrease.

However, a scattering-dependent quantity, i.e., $\sigma_{xx}$ has a large temperature dependence with values at 150 K and 300 K being $11.1 \times 10^4$ $\Omega^{-1}$cm$^{-1}$ and $3.8 \times 10^4$ $\Omega^{-1}$cm$^{-1}$, respectively [see Fig. 4(b)]. A small decrease in $\sigma_{xy}^A$ mentioned above can also be explained by examining the variation of the saturation magnetization $M_S$ in the same temperature range. This has been shown in the rightmost axis of Fig. 4(b), where we observe a change in the magnetization of about 25%. Moreover, both $\sigma_{xy}^A$ and $M_S$ show similar variation with temperature, suggesting a linear relationship between $\sigma_{xy}^A$ and $M_S$. This means that $\sigma_{xy}^A$ remains effectively constant, indicating the intrinsic origin of the AHE. This claim is further confirmed by analyzing the scaling behavior, where we plot $\rho_{yx}^A/M_S$ with respect to $\rho_{xx}^2$, resulting in a perfect straight line [Fig. 4(c)] [24]. This scaling law in other words suggests the same point that the anomalous Hall conductivity corrected by a varying saturation magnetization is a constant and thus does not depend on the scattering time. Although the $ab$-plane is the hard magnetic plane in the FM-1 phase, $M - \mu_0 H$ curves still saturate around 2 T magnetic field along [100]. We expect the same behavior in the corresponding Hall resistivity data when the magnetic field is applied along [010]. Figure 4(d) shows the Hall resistivity ($\rho_{zx}$) as a function of magnetic field where the current and magnetic field are applied along [100] and [010], respectively. The saturation behavior of $\rho_{zx} - \mu_0 H$ data in terms of applied magnetic field mimics that of the magnetization confirming the presence of the finite AHE. Interestingly, the maximum value of $\sigma_{xz}^A$ observed at 150 K is 456 $\Omega^{-1}$cm$^{-1}$, which is much higher than $\sigma_{xy}^A$ at the same temperature. However, the decrease of $\sigma_{xz}^A$ with increasing temperature is much faster [see Fig. 4(e)] than that of $\sigma_{xz}^A$; at 300 K, it could retain only 23% (105 $\Omega^{-1}$cm$^{-1}$) of the maximum value even though the total decrease in $M_S$ [right axis of Fig. 4(e)] is only 28% in the same temperature range. Furthermore, the temperature dependence of $M_S$ is not similar to that of $\sigma_{xz}^A$, suggesting a nonlinear relationship between $\sigma_{xz}^A$ and $M_S$. This suggests an extrinsic behavior of $\sigma_{xz}^A$. To confirm this, we plot $\rho_{yx}^A/M_S$ against $\rho_{xx}$ [see Fig. 4(f)], which fits well with a straight line confirming that $\sigma_{xz}^A$ is not temperature-independent and varies as a function of $\sigma_{xx}$ (or scattering time). Therefore, the AHE is dominated by the extrinsic skew scattering effect when the magnetic field is applied along the magnetically hard crystallographic [010] as compared to the completely intrinsic effect when the magnetic field is applied along the magnetically easy [001].

Now we turn our focus to the FM-2 phase where [100] or [010] is the magnetically easy axis compared to [001]. An important point to note here is that the [001] is very hard and does not show any signature of magnetic saturation in our magnetization measurements. Therefore, for the magnetotransport studies, we only consider the application of the magnetic field along [010] (easy axis). To our surprise, we did not observe any AHE at 5 K, even though the magnetization easily saturates at a small magnetic field of 0.4 T. The low temperature $\rho_{zx} - \mu_0 H$ data show a linear increase of $\rho_{zx}$ with increasing magnetic field [see Fig. 5(a)], indicating the normal Hall effect originating due to hole-dominated carrier transport. Only at temperatures of 50 K and above in this FM-2 phase do we see the feature of AHE in the $\rho_{zx} - \mu_0 H$ data. It is noteworthy that the corresponding $\sigma_{xz}^A$ is extremely large with a value as high as 1400 $\Omega^{-1}$cm$^{-1}$, but this value decreases very rapidly with increasing temperature, again indicating that it originates from an extrinsic origin [see Fig. 5(b)]. We perform the scaling analysis by plotting $\rho_{zx}^A/M_S$ against $\rho_{xx}$, and we find a perfect straight line fit [see Fig. 5(c)], confirming that it originates from the skew scattering mechanism.

In Fig. 6 we show the temperature dependence of the normal Hall coefficient $R_0$ in both the FM-1 [Fig. 6(a)] and FM-2 [Fig. 6(b)] phases. Although we observe a clear trend of decreasing $R_0$ with increasing temperature in the low temperature FM-2 phase, we do not see a clear trend of $R_0$ in the high-temperature FM-1 phase whether the field is applied along [001] or [010]. In this temperature range, the values of $R_0$ for the two different magnetic field directions are comparable within the error bar, which is expected since it depends only on the carrier concentration of the sample and not on the magnetic field direction.

In Fig. 7(a), we show $\rho_{yx} - \mu_0 H$ at different angles at 200 K (FM-1) with the magnetic field along [001] for $\theta = 0^0$ and along [100] for $\theta = 90^0$. A humplike feature can be observed in the $\rho_{yx} - \mu_0 H$ data, which is more pronounced at higher $\theta$'s. This feature is absent in the magnetization data [shown for $\theta = 75^0$ in Fig. 7(b)], which at first glance suggests that the feature in $\rho_{yx} - \mu_0 H$ could be a topological Hall signal. However, we argue that this feature is just a consequence of the nonorthogonal Hall geometry and not a topological Hall signal. It is well known that when the magnetic field is not orthogonal to the current, only the component of the field perpendicular to the current (in our case, the component along [001]) gives rise to an ordinary Hall effect (OHE), and the component parallel to the current does not contribute to the OHE. A similar argument applies to the AHE, where the role of the magnetic field is played by the magnetization of the sample. The magnetization shown in Fig. 7(b) is the net magnetization of the sample for the given geometry and not just a component of it. To investigate how the magnetization component along [001] ($M_z$) varies with the magnetic field, we refer to the domain theory of ferromagnetism. In a uniaxial ferromagnet, magnetic domains are typically aligned along the direction of the easy axis of magnetization ([001]). Initially, when no magnetic field is applied, there are two major domains of equal size [domain-A and domain-B in Fig. 7(c)] that are oppositely oriented parallel to [001]. To account for any other possible domains, we can consider two minor domains of equal size (domain-C and domain-D) that are oppositely oriented and orthogonal to the major domains. However, the following argument is equally valid even if we do not consider domains-C and -D. When no magnetic field is applied, the moments in the oppositely directed domains cancel each other out, resulting in zero net magnetization [see Fig. 7(c)]. When a magnetic field is applied at an angle $\theta$ away from [001], the domains that are inclined toward the magnetic field (domains-A and -D) will expand, while the domains that are inclined away from the magnetic field (domains-B and -C) will shrink in size via domain-wall motion. Because of the significant magnetocrystalline anisotropy of the sample, domain rotation is limited at lower fields. Domain-wall motion continues until domains B and C nearly disappear, resulting in a sharp increase in both the net magnetization ($M$) and the [001] component of the net magnetization ($M_z$). As the field is further increased, the domains begin to rotate toward the magnetic field so that $M$ continues to increase at a slower rate. Since the major magnetization is contributed by domain-A and it rotates away from [001] with the increasing magnetic field, $M_z$ begins to decrease. The domain rotation continues until all the domains are aligned with the magnetic field. $M$ and $M_z$ reach a saturation value at this point, where $M_z$ is given by the relationship $M_z = M \cos\theta$. The typical shape of the $M_z - \mu_0 H$ curve now looks like the shape of the $\rho_{yx} - \mu_0 H$ curve. Thus, we conclude that the humplike feature in the $\rho_{yx} - \mu_0 H$ data has a magnetic origin and is not related to the topological Hall effect. Since our argument is completely general to any anisotropic uniaxial ferromagnet, this humplike feature in $\rho_{yx} - \mu_0 H$ data in nonorthogonal Hall geometry should be common in such systems, which we indeed find in the literature [21,25].

The experimental result suggests that there is a change of the easy magnetic axis with temperature across FM-1 and FM-2 phases due to the change in interatomic distances, so we have performed electronic structure calculations for these two phases. These calculations suggest that the lower energy structures have a simple magnetic axis oriented along [001] at 300 K and along [100] at 0 K, when these two orientations are compared and these theoretical results are in good agreement with the experimental data. The dispersion of the energy bands at 300 K shows linear crossings at the Fermi energy ($E_F$) along Γ-X and Γ-N as shown in Fig. 8(a). Most of these degeneracies at $E_F$ are lifted in the presence of spin-orbit coupling (SOC) and form a nodal line. The red and blue lines represent the electronic bands with and without SOC. The main contribution to the bands near $E_F$ comes from the $d$-orbital of Mn. We have also calculated the $z$-component of the Berry curvature (BC) along the high-symmetry path as shown in Fig. 8(b), and we note distinct BC peaks along Γ-X and Γ-N. To further analyze the class of degeneracies, the energy gap is calculated in the presence of SOC. SmMn$_2$Ge$_2$ belongs to the *I4/mmm* space group, which has three mutually perpendicular mirror planes, giving rise to nodal lines in the absence of SOC [see Fig. S4(a)]. However, the ferromagnetic ordering along [001] and the finite SOC break the mirror symmetry of $k_y = 0$ and $k_x = 0$. Consequently, the degeneracy of the nodal lines is lifted as shown in Fig. S4(b). This leads to significant Berry curvature contributions along the nodal line, as shown in Fig. 8(c), which contribute directly to the intrinsic component of the AHC. The variation of the AHC with energy is shown in Fig. 8(d). We found an intrinsic AHC ($\sigma_{xy}^{int}$) at $E_F$ of about 100 Ω$^{-1}$cm$^{-1}$, which is in excellent agreement with the pure intrinsic AHC observed in experiments in the FM-1 phase. The structure at 0 K favors the magnetic ordering along [100], and we analyze this system with the same mirror symmetries. The electronic band structure is shown in Fig. 8(e), and we observe that the band crossing points have shifted by 0.107 eV above $E_F$. These crossings remain intact in the presence of SOC and magnetic ordering along [100]. The red and blue lines represent the energy bands in the presence and absence of SOC. The BC at $E_F$ is shown in Fig. 8(f) and is negligible. The shift of these degenerate points above $E_F$ leads to a lack of intrinsic AHC because there is no BC contribution from bands above $E_F$, which is consistent with the experimental observations in the FM-2 and FM-1 phases.

**III. DISCUSSION**

In this paper, we show that the nature of the AHE can be changed from completely intrinsic to extrinsic by changing the direction of the magnetic field from one crystallographic axis to another. This is even more remarkable since in SmMn$_2$Ge$_2$, these two crystallographic axes [001] and [100] can both be magnetically easy axes depending on the temperature range chosen. We have claimed the complete intrinsic AHE in the FM-1 phase with the magnetic field along [001] based on the perfect linear fit between $\rho_{yx}^A/M_S$ and $\rho_{xx}^2$ in Fig. 4(c). Although this is sufficient to confirm the intrinsic nature of the AHE, one can also calculate the intrinsic contribution from the slope of the linear fit. Using the value of $M_S$ at 150 K, we obtain the intrinsic component $\sigma_{xy}^{A(int)} = 107\ \Omega^{-1}cm^{-1}$ demonstrating the absence of any extrinsic component. However, when applying the magnetic field along [010], regardless of whether we are in FM-1 ([010] is the hard axis) or FM-2 ([010] is the easy axis) phase, we observe only an extrinsic skew scattering dominated AHE. We have plotted $\sigma_{Hall}^A$ versus $\sigma_{xx}$ in Fig. 9 for our SmMn$_2$Ge$_2$ along with other well-known AHE systems from the literature [5,7,26–34]. It is clearly seen that for magnetic field along [010], $\sigma_{Hall}^A$ increases monotonically as a function of $\sigma_{xx}$ across the FM-1 and FM-2 phases (see shaded region in Fig. 9). This is expected because the skew scattering mediated AHC depends only on the scattering time $\tau$ (or $\sigma_{xx}$) [27,35] provided that $M_S$ does not have large fluctuations, which is true for SmMn$_2$Ge$_2$ over the FM-1 and FM-2 phases. One can also define a skewness parameter $\left(S_{skew} = \frac{\sigma_{Hall}^A}{\sigma_{xx}}\right)$ that depends only on the type of impurity [2]. Since the type of impurity is expected to be the same for SmMn$_2$Ge$_2$ over FM-1 and FM-2 phases, we see a monotonic increase in $\sigma_{Hall}^A$ with respect to $\sigma_{xx}$ (shaded region in Fig. 9). The surprising result that at 5 K we observe only the normal Hall effect [see Fig. 5(a)] in a ferromagnetic state with the magnetic field along the easy magnetic axis can be understood based on how the normal Hall conductivity $\left(\sigma_{xy}^N\right)$ and the skew scattering mediated anomalous Hall conductivity $\left(\sigma_{xy}^{A,skew}\right)$ depend on $\tau$. Since the $RRR$ of SmMn$_2$Ge$_2$ crystals is of the order of $10^2$, the value of $\tau$ is very large at low temperature. $\sigma_{xz}^N$ and $\sigma_{xz}^{A,skew}$ vary as $\tau^2$ and $\tau$, respectively. Hence, $\sigma_{xz}^N$ dominates over $\sigma_{xz}^{A,skew}$ at low temperature [36]. As the temperature is increased, the value of $\tau$ decreases rapidly, and thus we see a clear signature of AHE from 50 K onwards.

**CONCLUSION**

We have investigated the anisotropic magnetotransport properties of SmMn$_2$Ge$_2$, which undergoes several magnetic phase transitions. In the high-temperature ferromagnetic phase, we show that the observed anomalous Hall effect is completely intrinsic when the magnetic field is applied along the magnetically easy *c*-axis. This can be completely tuned to an extrinsic anomalous Hall effect mediated by skew scattering when the magnetic field direction is turned to the magnetically hard *a*-axis. In the low-temperature ferromagnetic phase, the *a*-axis is a magnetically easy axis, yet we do not observe any anomalous Hall effect at 5 K because the normal Hall effect easily dominates due to the large scattering time. Upon increasing the temperature, we observe an extremely large anomalous Hall conductivity (1400 $\Omega^{-1}$cm$^{-1}$), which originates form the skew scattering mechanism. In support of our experimental findings, the density functional theory-based calculations show that when the spins are aligned along [001], gapped nodal lines near the Fermi energy result in Berry curvature induced anomalous Hall conductivity of ~100 $\Omega^{-1}$cm$^{-1}$. When the spins are aligned along [100], these nodal line features shift above the Fermi energy, producing a negligible anomalous Hall conductivity.

**METHOD**

High-quality single crystals of SmMn$_2$Ge$_2$ were prepared by the flux method using indium (In) flux. Pieces of the elements were mixed in the stoichiometric ratio of Sm : Mn : Ge : In = 1 : 2 : 2 : 60 and placed in an alumina crucible that was sealed under vacuum. The sealed quartz tube was placed in the furnace, slowly heated to 1050 ºC, and held for 24 h. It was then slowly cooled to 700 ºC at a rate of 2 ºC/h. To separate single crystals from In flux, the tube was taken out at 700 ºC and centrifuged. Platelike single crystals were obtained. The crystal structure of prepared single crystals was characterized using an x-ray diffractometer (XRD, smart lab, Rigaku) equipped with Cu-$K\alpha$ radiation. Energy-dispersive x-ray spectroscopy (EDX) was performed to verify the chemical composition of the samples. Magnetic measurements were performed using a vibrating sample magnetometer (VSM) on a physical properties measurement system (PPMS, Dynacool, Quantum Design). The electrical transport measurements under magnetic field were performed using the electrical transport option (ETO) of the PPMS. A rectangular crystal of 1.2 mm× 0.59 mm × 0.54 mm was used for magnetotransport and magnetic measurements. A standard four-probe method was used for longitudinal and Hall resistivity measurements. To cancel the longitudinal resistivity contribution due to probe misalignment, the Hall resistivity was antisymmetrized using the relationship $\rho_H(H) = \frac{\rho_H(+H) - \rho_H(-H)}{2}$.

*Ab initio* calculations of the electronic structure of SmMn$_2$Ge$_2$ were performed using density functional theory (DFT) with the generalized gradient approximation (GGA) of the exchange correlation functional [37] implemented in the Vienna ab initio simulation package code [38]. The effective Coulomb exchange interaction $U_{\text{eff}}$ ($U-J$), where $U$ and $J$ are the Coulomb and spin exchange parameters, was used to incorporate the strong on-site Coulomb repulsion, also called Hubbard $U_{\text{eff}}$ = 3.0 eV for the *d*-orbitals of the Mn atoms. For this value of $U$, the calculated value of the average magnetic moment is in agreement with the experiment. The cutoff energy for the expansion of the wave functions into the plane-wave basis was kept constant throughout at 600 eV. The Brillouin zone was sampled in the *k*-space of the Monkhorst-Pack scheme for the calculation. The equilibrium structure served as the basis for the $k$-grid, which was 10 × 10 × 6. In the next step, Wannier functions were extracted from the DFT band structure using the wannier90 package [39,40]. A tight-binding Hamiltonian was constructed from the Wannier functions to calculate the Berry curvature via the Kubo formula using wanniertools [41]. The intrinsic Hall conductivity ($\sigma^{int}_{xy}$) was calculated by integrating the $z$-component of the Berry curvature ($\Omega_z$) over all occupied states in the entire Brillouin zone, taking into account the spin-orbit coupling.

*Note added.* Recently, we became aware of a study on magnetotransport of SmMn$_2$Ge$_2$ [42].

## ACKNOWLEDGMENTS

N.K. acknowledges DST for financial support through Grant Sanction No. CRG/2021/002747 and Max Planck Society for funding under Max Planck-India partner group project. This research project made use of the instrumentation facility provided by the Technical Research Centre (TRC) at the S. N. Bose National Centre for Basic Sciences, under the Department of Science and Technology, Government of India. M.K. acknowledges DST for funding through Grant No. CRG/2020/000754. J.S. thanks UGC for financial support.

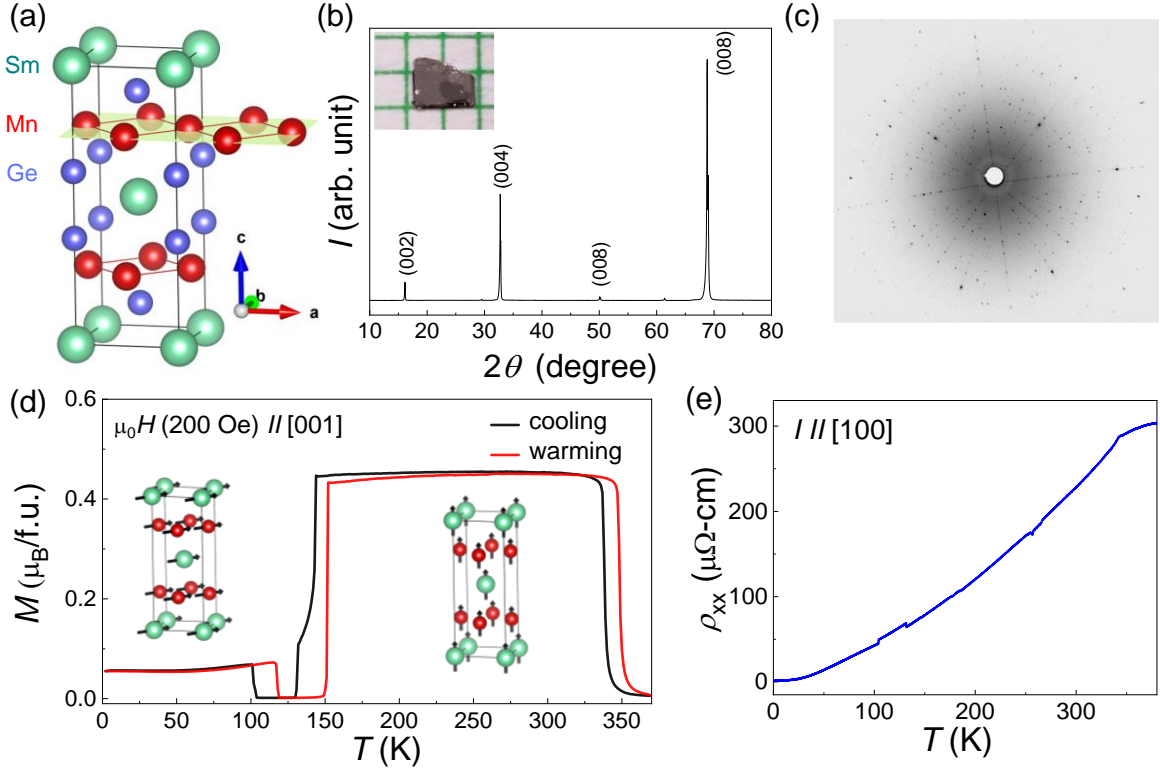

**FIG. 1.** (a) A unit cell of SmMn$_2$Ge$_2$ with square nets (highlighted by a yellow plane) of Mn atoms. (b) x-ray diffraction pattern of SmMn$_2$Ge$_2$ single crystal with only the *ab*-plane exposed to the x-ray. The diffraction pattern shows only (00*l*) reflections. The inset shows a typical single crystal placed on a graph paper. (c) The Laue diffraction pattern obtained by exposing the single crystal with the x-ray beam along [00*l*]. (d) Magnetization vs temperature data obtained at a magnetic field of 200 Oe along [001] in the field-cooled condition. Black and red lines show the data while cooling and warming, respectively. Spin alignments are shown for the high-temperature ferromagnetic phase (FM-1) and the low-temperature ferromagnetic phase (FM-2). (e) Longitudinal resistivity as a function of temperature in the absence of the magnetic field by applying a current along [100].

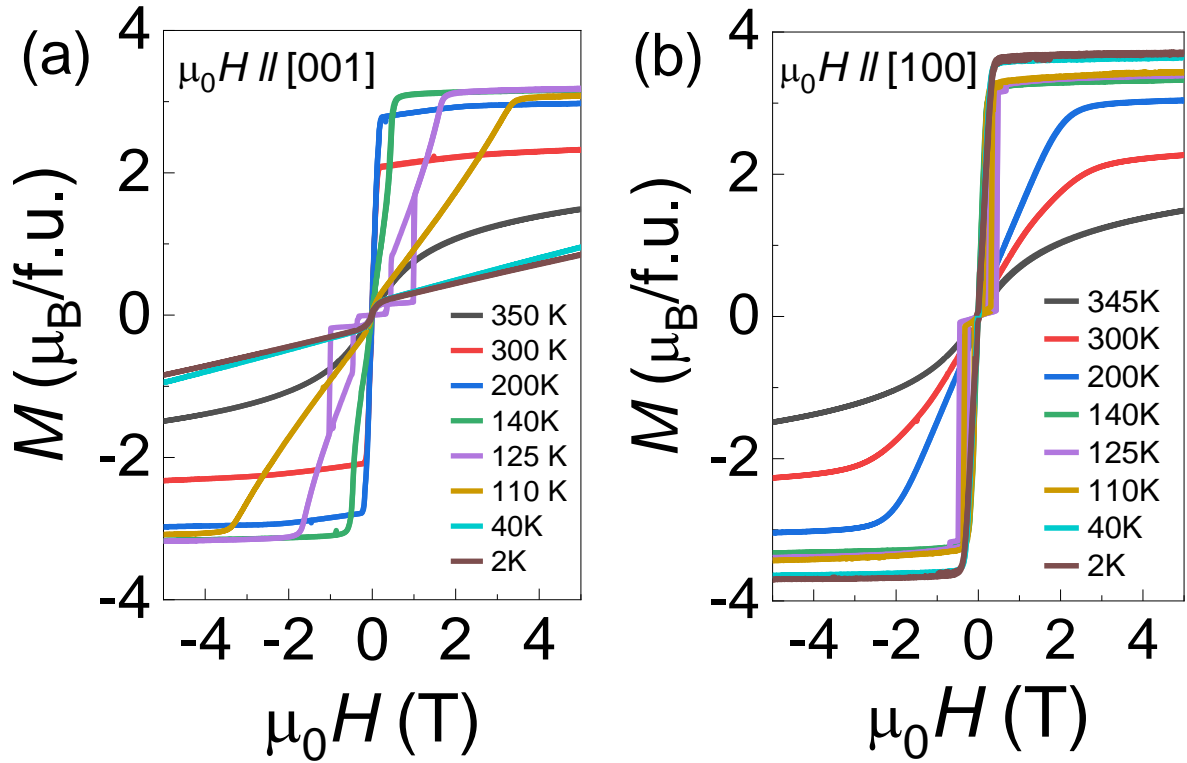

**FIG. 2.** Magnetization as a function of magnetic field at various temperatures across FM-1, AFM, and FM-2 phases for the applied magnetic field along (a) [001] and (b) [100].

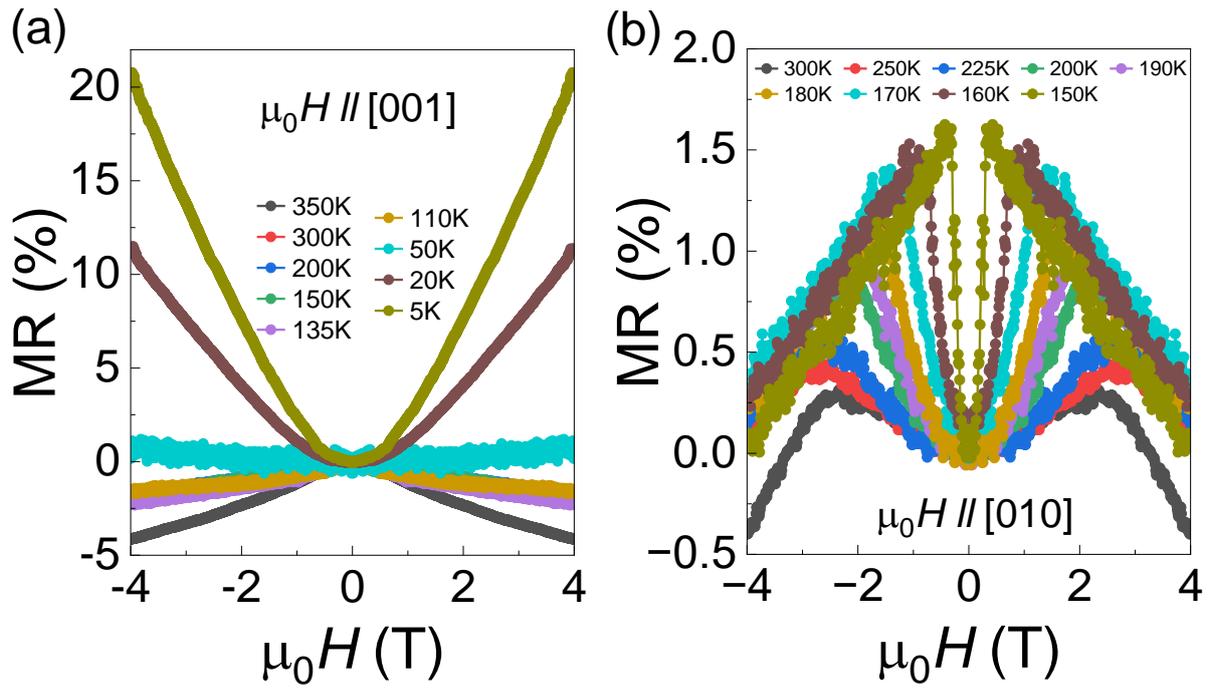

**FIG. 3.** Magnetoresistance data for various temperatures with the magnetic field applied along (a) [001] and (b) [010].

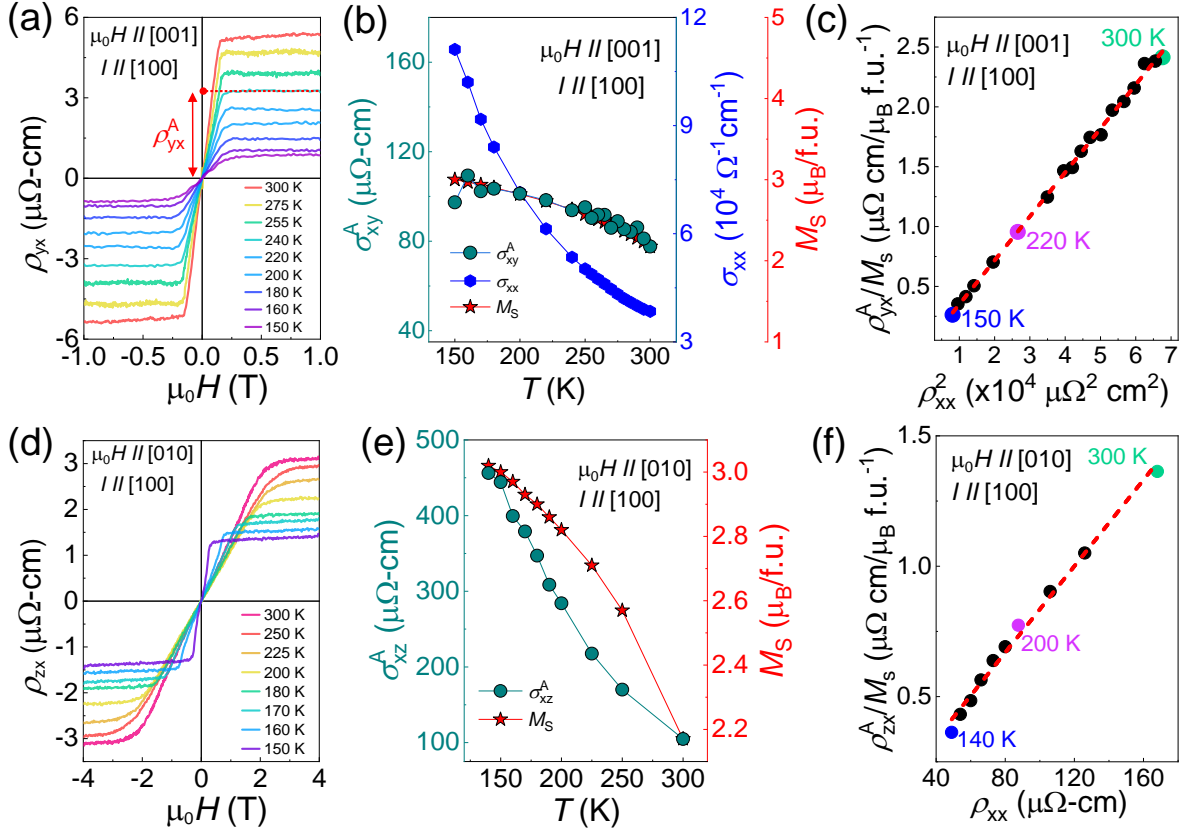

**FIG. 4.** (a) $\rho_{yx} - \mu_0 H$ data for $\mu_0 H$ along [001] at temperatures in the FM-1 phase. (b) Corresponding $\sigma^A_{xy}$ (left axis), $\sigma_{xx}$ (right axis), and $M_S$ (rightmost axis) with temperature in the FM-1 phase. (c) $\rho^A_{yx}/M_S$ vs $\rho^2_{xx}$ for $\mu_0 H$ along [001] in the FM-1 phase. (d) $\rho_{zx} - \mu_0 H$ data for $\mu_0 H$ along [010] at temperatures in the FM-1 phase. (e) Corresponding $\sigma^A_{xz}$ (left axis) and $M_S$ (right axis) with temperature in the FM-1 phase. (f) $\rho^A_{zx}/M_S$ vs $\rho_{xx}$ for $\mu_0 H$ along [010] in the FM-1 phase.

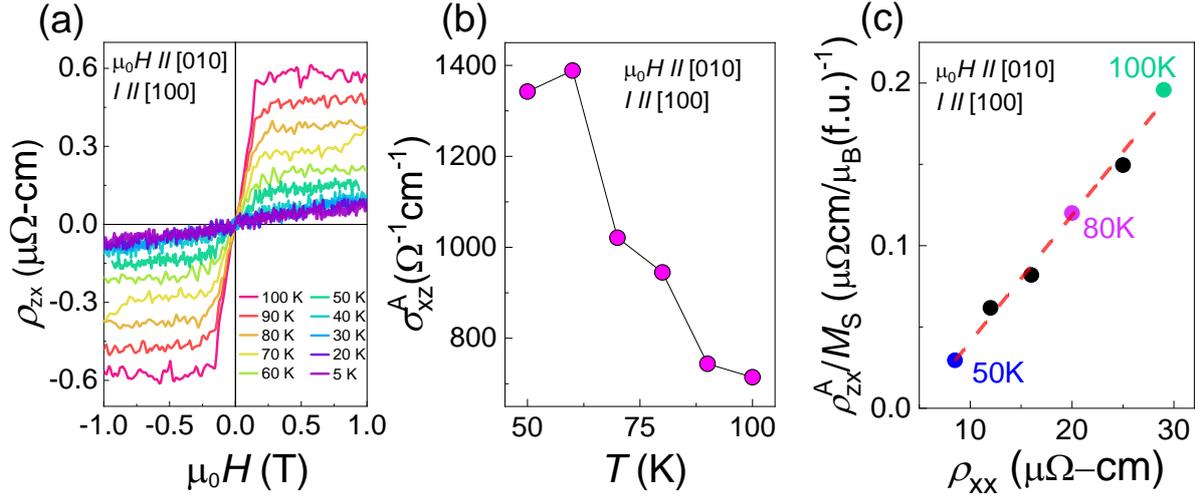

**FIG. 5.** (a) $\rho_{zx} - \mu_0 H$ data for $\mu_0 H$ along [010] from 5 to 100 K in the FM-2 phase. (b) $\sigma^A_{xz}$ vs $T$ data for $\mu_0 H$ along [010] in the FM-2 phase. (c) $\rho^A_{zx}/M_S$ vs $\rho_{xx}$ for $\mu_0 H$ along [010] in the FM-2 phase.

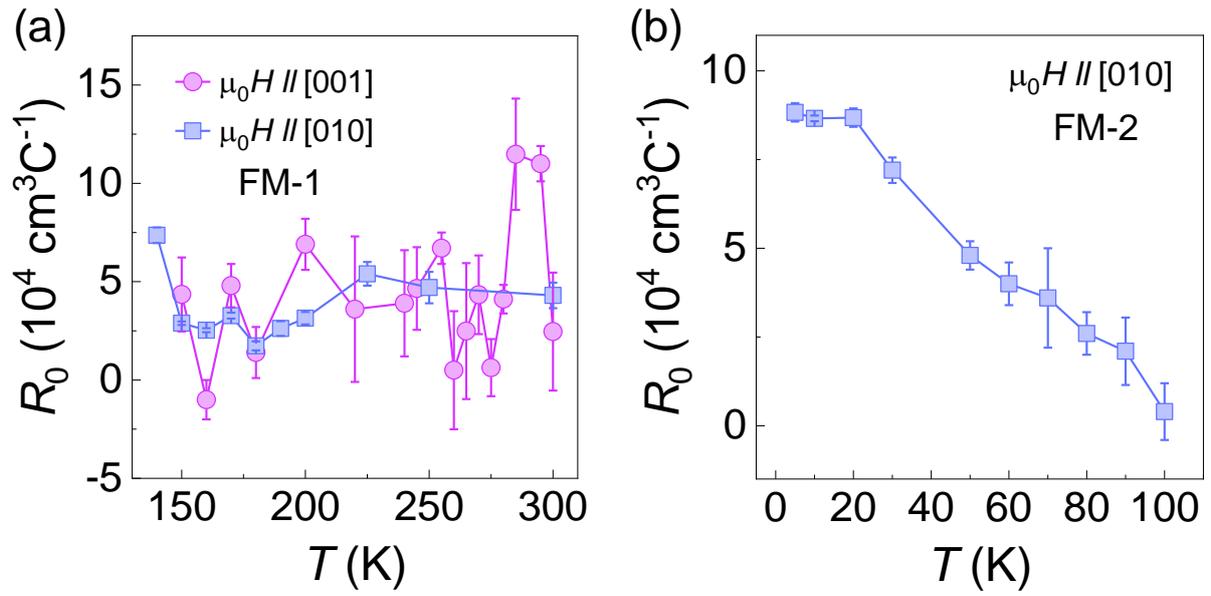

**FIG. 6.** Dependence of normal Hall coefficient $R_0$ with temperature in (a) the FM-1 phase and (b) the FM-2 phase.

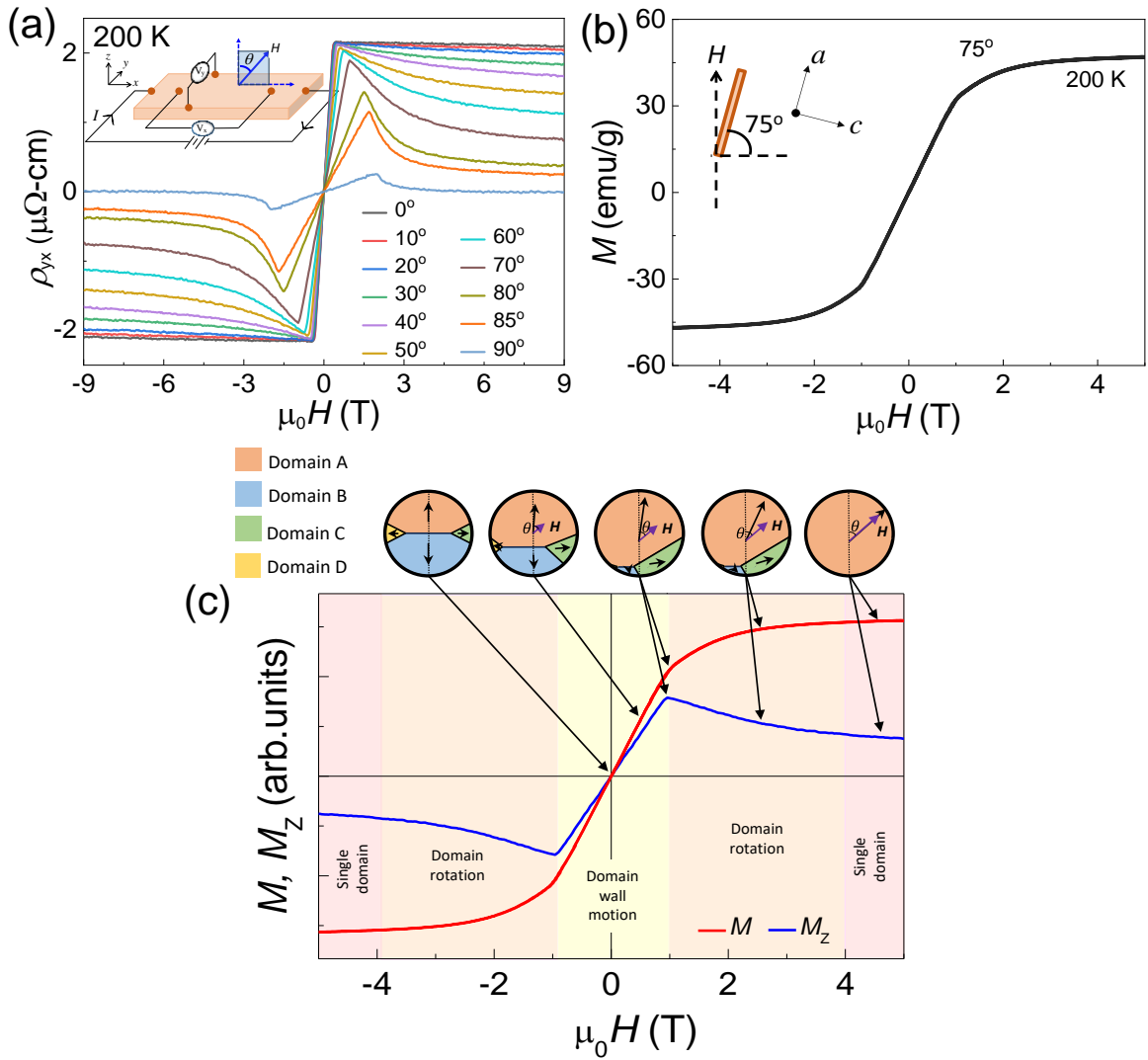

**FIG. 7.** (a) $\rho_{yx} - \mu_0 H$ data at 200 K at different angles with $\theta = 0^0$ and $90^0$ corresponding to $\mu_0 H$ along [001] and [100], respectively. (b) Corresponding $M$ vs $H$ data at $\theta = 75^0$. (c) Net magnetization ($M$) and the $z$-component of the net magnetization ($M_z$) as a function of magnetic field applied at an angle $\theta$ away from the $c$-axis as explained by the possible domain evolution.

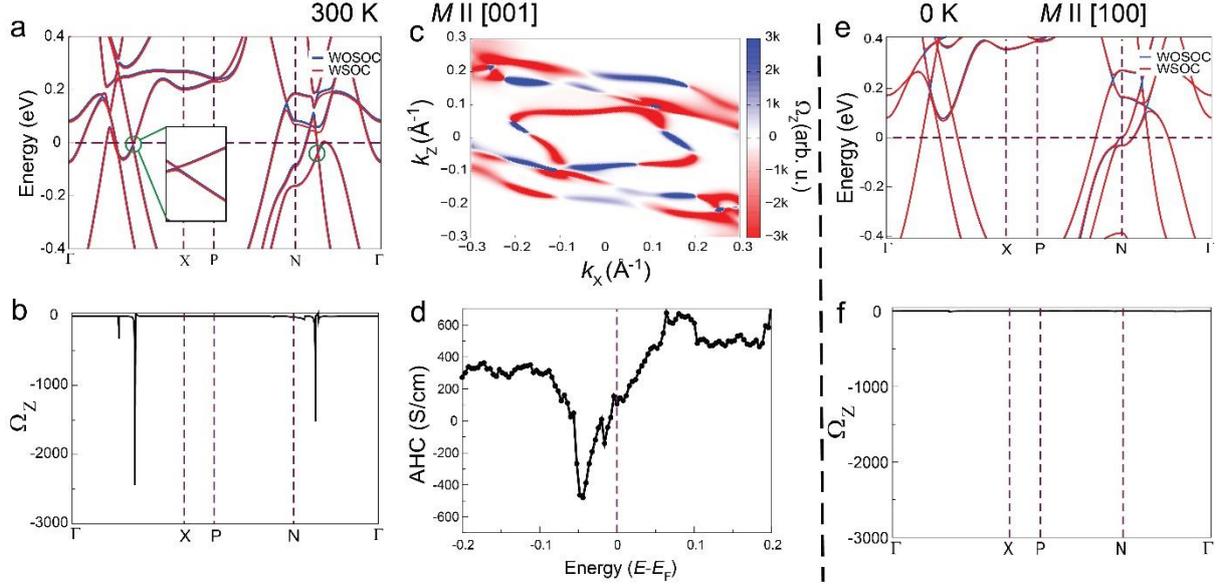

**FIG. 8.** (a) The band structure of SmMn$_2$Ge$_2$ without SOC and with SOC at 300 K where the spins are oriented along [001]. The gapped nodal lines are shown in the inset. (b) The $z$-component of Berry curvature peaks along the high-symmetry lines due to the gapped nodal lines. (c) Along the gapped nodal line, the Berry curvature distribution is shown at the $k_y = 0$ plane. (d) Energy ($E$-$E_F$) dependence of the AHC for SmMn$_2$Ge$_2$. (e) The band structure of SmMn$_2$Ge$_2$ without SOC and with SOC at 0 K. (f) The $z$-component of Berry curvature along the high symmetry lines.

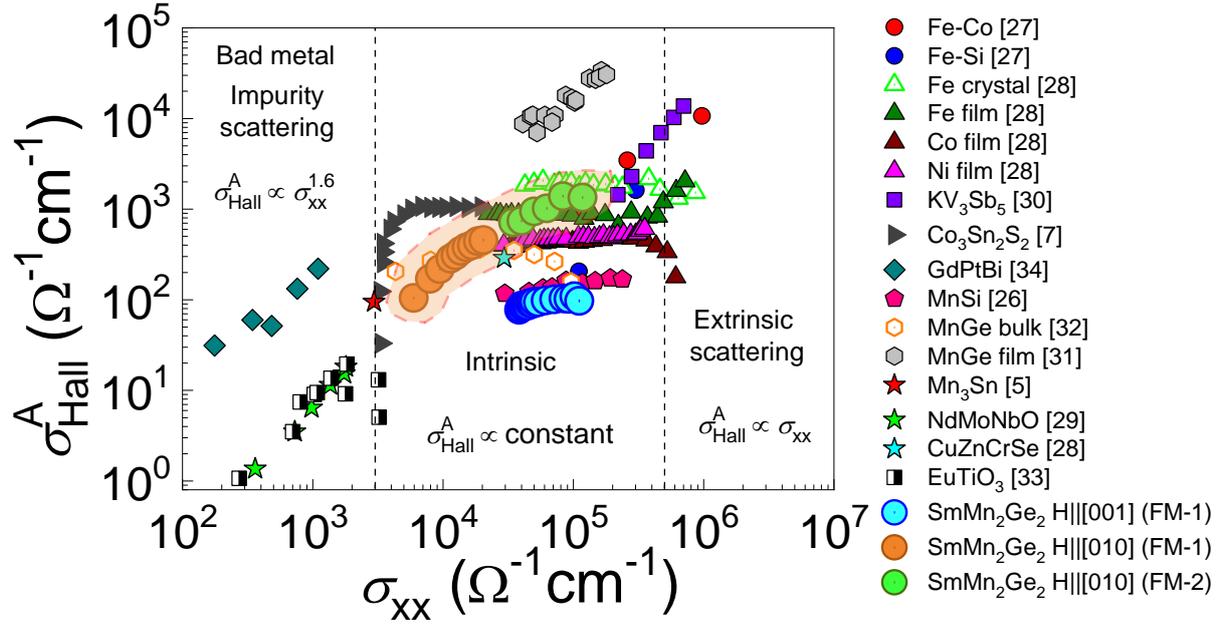

**FIG. 9.** $\rho^A_{Hall}$ vs $\sigma_{xx}$ for various materials showing AHE along with the data of SmMn$_2$Ge$_2$ in the FM-1 phase with $\mu_0 H$ along [001] and [010] and in the FM-2 phase with $\mu_0 H$ along [010]. The shaded region shows the extrinsic skew scattering mediated AHE in SmMn$_2$Ge$_2$.

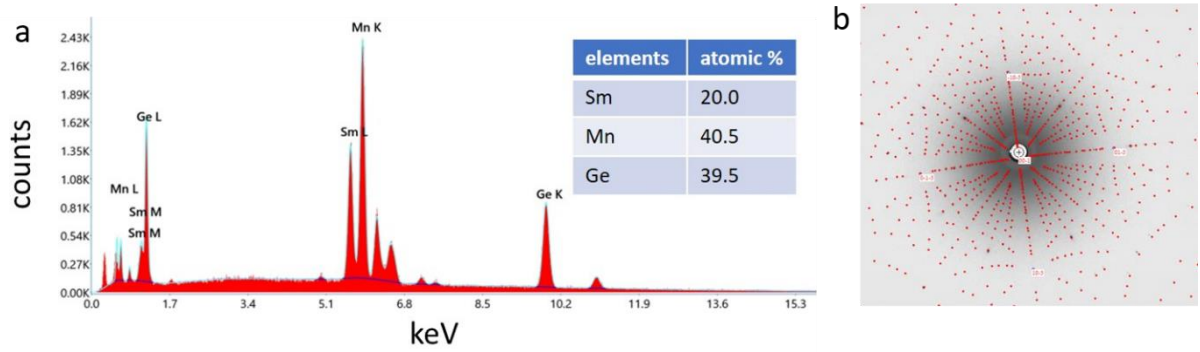

**FIG. S1.** (a) Energy dispersive X-ray spectrum of a typical SmMn$_2$Ge$_2$ single crystal. The inset shows the corresponding composition obtained from the fitting of the spectrum. (b) Fitting of the x-ray diffraction pattern (red spots) with the space group *I*4/*mmm*.

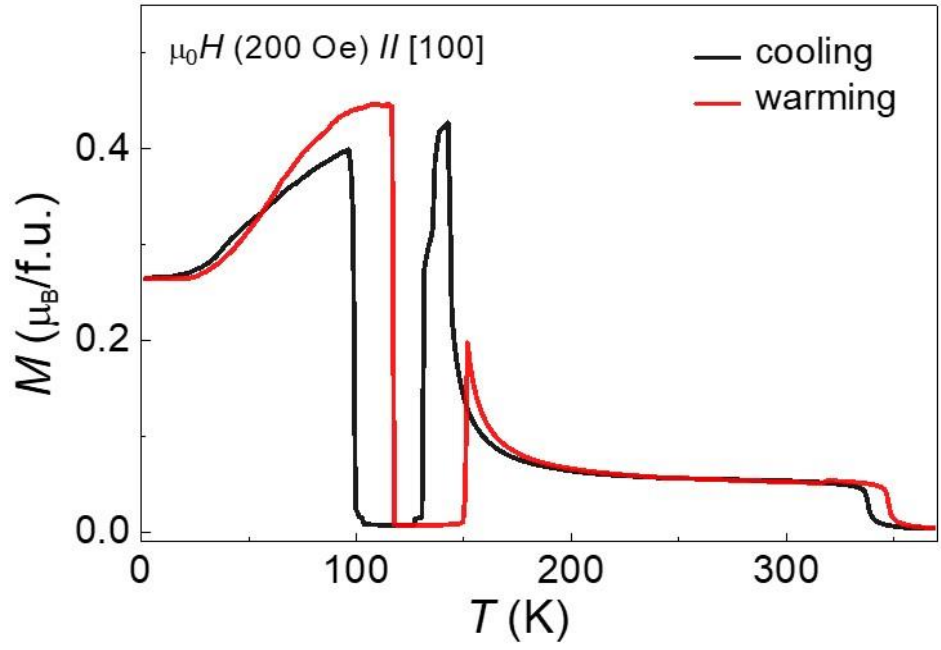

**FIG. S2.** Magnetization versus temperature data obtained at a magnetic field of 200 Oe along [100] in the field cooled condition. Black and red lines show the data while cooling and warming, respectively.

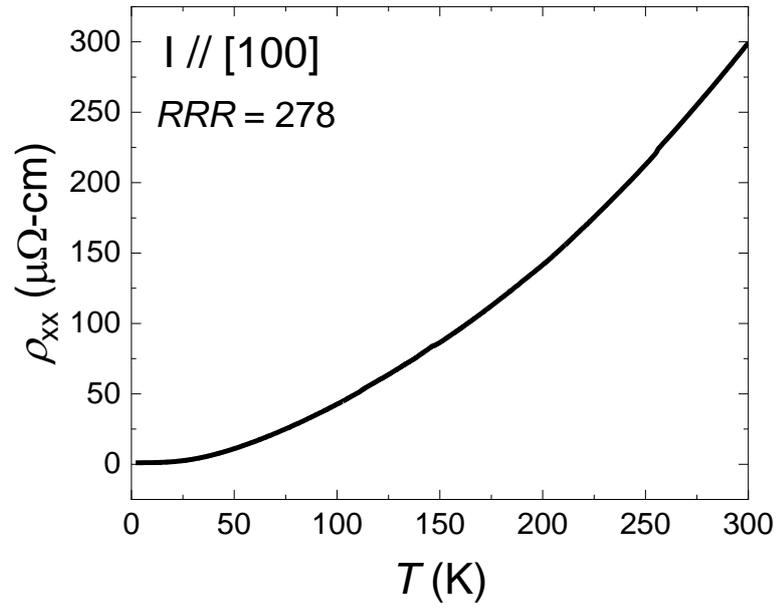

**FIG. S3.** Longitudinal resistivity as a function of temperature in the absence of magnetic field by applying a current along [100] on a crystal other than shown in the main text. The corresponding value of the *RRR* is 278.

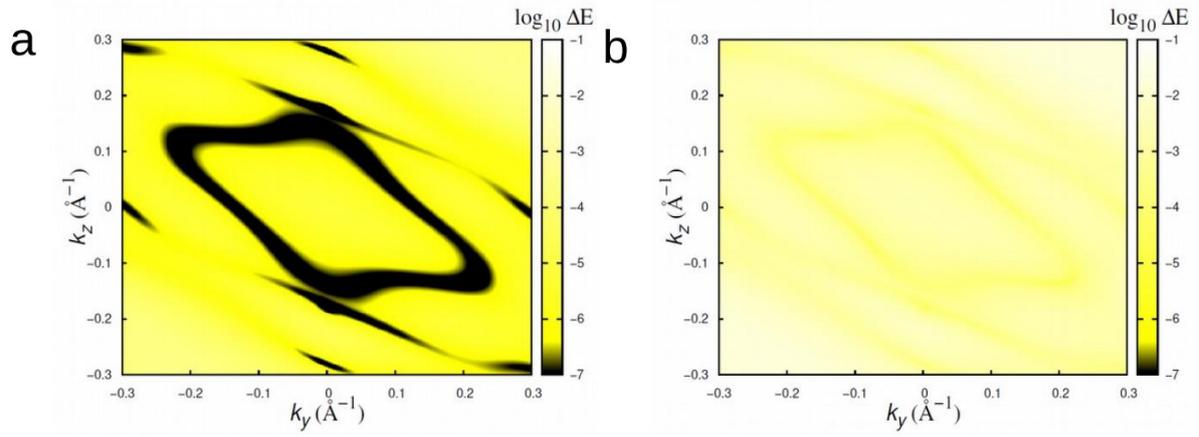

**FIG. S4.** (a) Energy gap between two intersecting bands in the vicinity of $E_\text{F}$ on the $k_\text{x}$-$k_\text{z}$ plane, where the black oval closed line denotes the nodal line. (b) The nodal line is gapped out in the $k_\text{x}$-$k_\text{z}$ plane in the presence of SOC, and the gap is represented by a white line.